# On Estimation of Discretization Error Norm for Compressible Euler Equations

A.K. Alekseev[1*], A.E.Bondarev[2], I. M. Navon[3]

[1]Moscow Institute of Physics and Technology, Moscow, Russia

[2]Keldysh Institute of Applied Mathematics RAS, Moscow, Russia

[3]Department of Scientific Computing, Florida State University, Tallahassee, FL 32306-4120, USA

## Abstract

The issue of single-grid discretization error estimator, operating in the postprocessor mode, is addressed. An ensemble of numerical solutions, obtained using solvers of different order of accuracy, is shown to provide an upper estimate for the norm of the discretization error. This operation is feasible, if this ensemble contains separate clusters of "accurate" and "inaccurate" solutions. Numerical tests for supersonic flows, governed by two dimensional Euler equations are presented and compared with analytical solutions. These tests demonstrate potentials and restrictions of the ensemble based error norm estimation.

## 1. Introduction

The standard grid convergence strategy is based on the heuristic rule by C. Runge [1]. From this perspective, if the difference between two approximate solutions on coarse grid and on the fine grid is small, then solutions are close to exact one. However, from a practical needs perspective one should desire a quantitative estimate of the form $\|u_h - \tilde{u}\| \le \delta$ with computable $\delta$. Formally, the Richardson method [2] is close to this ideal. It enables us to determine the refined solution and the error estimate using a set of solutions computed on different meshes, which should belong to the asymptotic range of convergence. Two meshes are necessary if a single error order exists in flow field. Unfortunately, in most CFD problems the error order on different flow structures varies, so the order should be determined additionally, requiring at least three consequent meshes and causing an ill-posed statement. The present paper addresses an alternative to the Richardson method. The set of solutions is collected on the same mesh using different solvers that provide an estimate of the global error norm. Calculations may be terminated if a preassigned error level $\|u_h - \tilde{u}\| \le \delta$ is attained.

At present, CFD uses a wide selection of numerical methods that are characterized by a rich variety of properties such as monotonicity, conservativity, order of approximation etc. This is naturally caused by the search for more "accurate" numerical solutions. The abundance of numerical methods may provide some additional opportunities for quantitative analysis of CFD results, which we consider herein.

The "accurate" and "inaccurate" numerical schemes are often compared in terms such as the truncation error and the discretization error.

The truncation error $\delta u$ is obtained via Taylor series decomposition of the discrete operator $A_h u_h = f_h$, which approximates a system of PDE, formally denoted herein as $Au = f$. The truncation error dependence on the spatial step $h$ is usually presented as $\delta u = O(h^n)$, where the order $n$ is equal to the minor order of series terms. At the next stage of analysis, the approximation error $\Delta u = u_h - u$ (of real practical interest) should be estimated. The approximation error may be described by the tangent linear equation $A\Delta u = \delta u$ whose formal solution is $\Delta u = A^{-1}\delta u$.

*corresponding author

Email addresses: [1]alekseev.ak@phystech.edu, [2]a_bond2001@mail.ru, [3]navonim@gmail.com

For linear problems, the approximation error $\|\Delta u\| = O(h^n)$ tends to zero with the same order $n$ (by Lax theorem, [3]), if the discrete operator is well-posed (i.e. the inverse operator is uniformly bounded $\|A_h^{-1}\| < C$).

The estimation of the error order is significantly more complicated for the case of nonlinear equations with discontinuities [4, 5,6, and 7]. In this event, the discretization error comprises the components of different orders, which occur at various elements of the flow structure (such as shock waves or expansion fans). Thus, the observed order of local convergence may not be equal to the nominal order of the approximation error even for the asymptotic range.

There are several general directions for the error estimation. *A priori* error estimation is the most widely used approach for the error analysis and may be expressed in the form $\|\Delta u\| < C \cdot h^n$, which contains the unknown constant independent of the current numerical solution. It is the theoretical basis both for the development of numerical algorithms and for the mesh refinement strategy commonly used in CFD. *A posteriori* error estimation [1,8] has the form $\|\Delta u\| \le C_h e_h$, where $C_h$ is the computable stability constant, which depends on the numerical solution, and $e_h$ is the computable indicator of the truncation error. At present, the main successes in this direction are achieved for elliptic partial differential equations and finite element methods. In most of practical applications the stability constant is not estimated, while the error indicator is used for mesh adaptation.

The truncation error $\delta u$ estimates may serve as the simplest computable error indicator. It may be computed by the action of the high order scheme stencil on the precomputed flow field [9, 10], by the action of the differential operator on the interpolation of the numerical solution [11] or via the differential approximation [12, 13].

The application of the truncation error $\delta u$ implies the calculation of the discretization (global) error $\Delta u = A^{-1}\delta u$. A survey of the error calculation methods may be found in [14]. In the simplest form, the estimation of this error may be performed using a defect correction [9, 15]. In the defect correction frame, the truncation error $\delta u$ is used as the source term inserted in the discrete algorithm in order to correct the solution. However, the total subtraction of the error implies the elimination of the scheme viscosity that may cause oscillations near discontinuities or an activation of some additional dissipation sources, which engenders their own error. Also, the estimation of the error may be performed via a linearized problem [16], complex differentiation [17] or by adjoint equations [10,11,13,18]. Usually, adjoint equations are applied to the estimation of the uncertainty of certain valuable functional (drag, lift coefficients etc.). Nevertheless, the approach of [13] enables one to estimate the norm of the solution error. Unfortunately, it requires solving a number of adjoint problems, which is proportional to the number of grid nodes that implies an extremely high computational burden.

The presence of unknown components of the truncation error is the general disadvantage of above discussed residual-based error estimation methods. The differential approximation based methods use minor terms of Taylor series [13] and do not account for remaining higher terms. The postprocessor based methods do not account for the higher scheme truncation errors [10] or the interpolation errors [11].

The present paper considers the feasibility of finding the discretization error norm using the ensemble of calculations performed by solvers of different approximation order on the same mesh. We shall refer to this operation as the "ensemble based error estimation". Since the analysis is conducted in the space of numerical solutions, the truncation error is accounted for implicitly and completely. It is important that a mesh refinement is not used, thus requiring only moderate computational costs.

$L_1$ norm seems to be natural for problems dealing with shocks, since most results on approximation error are obtained in this norm [4,5]. However, most practical research interests are related to valuable functionals (lift, drag, etc). Their uncertainty may be related to the $L_2$ error norm via the Cauchy–Bunyakovsky–Schwarz inequality. For this reason, the discrete $L_2$-equivalent norm is used herein, while some results for other norms may be found in [19].

The presentation of this paper is organized as follows. In Section 2 we discuss the estimation of discretization error norm based on *a priori* information regarding error magnitude rating. Section 3 considers *a posteriori* analysis of error norm relations provided by the ensemble of numerical solutions obtained by different solvers. The supersonic shocked flows, described by two dimensional Euler equations are considered as the test problems in Section 4. The results of the ensemble based error norm estimation in comparison with the true error are presented for a set of solvers. The Section 5 provides a discussion concerning the features of considered approach to the error analysis. Conclusions are presented in the Section 6.

## 2. The estimate of discretization error norm via the set of approximate solutions

Let's consider the ensemble of numerical solutions obtained using finite difference or finite volume schemes of different accuracy orders on the same grid. Let the relation of the approximation error of these schemes be known *a priori*.

We denote the numerical solution as the vector $u^{(i)} \in R^N$ ( $i$ is the scheme number, $N$ is the number of grid points respectively), the values of unknown exact solution at nodes of this grid (further denoted as exact solution) as $\tilde{u} \in R^N$ and use a discrete $L_2$-equivalent norm. The unknown deviation of exact solution values at grid points $\tilde{u} \in R^N$ from the computed solution is assuming the form $\left\| u^{(i)} - \tilde{u} \right\|_{L_2} = r_i$. The numerical solutions $u^{(i)}$ are located at surfaces of concentric hyperspheres with the centre at $\tilde{u}$ and radii $r_i$ (unknown).

In the simplest event of two numerical solutions $u^{(1)}$ and $u^{(2)}$ with *a priori* errors relation $r_1 \geq 2 \cdot r_2$ ( $r_1 = \left\| \tilde{u} - u^{(1)} \right\|_{L_2}, r_2 = \left\| \tilde{u} - u^{(2)} \right\|_{L_2}$ ) the following theorem may be stated.

**Theorem 1.** *Let the norm of difference of two numerical solutions* $u^{(1)} \in R^N$ *and* $u^{(2)} \in R^N$

$$\left\| du_{1,2} \right\|_{L_2} = \left\| u^{(1)} - u^{(2)} \right\|_{L_2} \qquad (1)$$

*be known from computations and there is the a priori information*

$$\left\| \tilde{u} - u^{(1)} \right\|_{L_2} \geq 2 \cdot \left\| \tilde{u} - u^{(2)} \right\|_{L_2}, \qquad (2)$$

*then the norm of approximate solution* $u^{(2)}$ *error is less than the norm of difference of solutions:*

$$\left\| \tilde{u} - u^{(2)} \right\|_{L_2} \leq \left\| du_{1,2} \right\|_{L_2} \qquad (3)$$

**Proof.** The triangle inequality [20] for $u^{(1)}$, $u^{(2)}$, and $\tilde{u}$ assumes the form $r_1 \leq r_2 + \left\| du_{1,2} \right\|_{L_2}$, which is equivalent to $r_1 - r_2 \leq \left\| du_{1,2} \right\|_{L_2}$. By accounting (2) as $r_1 - r_2 \geq r_2$ one obtains $r_2 \leq r_1 - r_2 \leq \left\| du_{1,2} \right\|_{L_2}$ and, finally, the desired expression (Eq. 3): $r_2 \leq \left\| du_{1,2} \right\|_{L_2}$.

## 3. A posteriori analysis of discretization error norm rating

The widespread opinion that the schemes of higher order are more accurate has an asymptotic origin and, usually, is not supported by quantitative error norm estimates. So, the evident weakness of *Theorem 1* from the standpoint of applications is the assumption of the existence of solutions with *a priori* ranged error.

Herein, we consider *a posteriori* check of error ranging. The collection of distances between solutions $\left\| du_{i,j} \right\|_{L_2}$ (norms of difference of numerical solutions) enables a detection of nearby and distant solutions.

Let us compare subsets of distances $\left\|du_{1,j}\right\|_{L_2}$ and $\left\|du_{i,j}\right\|_{L_2}$, $(i \neq 1)$, where $u^{(1)}$ denotes the maximally incorrect solution. If $r_1 >> r_i$, the total set of distances $\left\|du_{i,j}\right\|_{L_2}$ is split into a subset with great values $\left\|du_{1,j}\right\|_{L_2}$ (distances from accurate solutions to inaccurate one) and a subset of distances between more accurate solutions $\left\|du_{i,j}\right\|_{L_2}$. This is caused by the asymptotics $\left\|du_{1,j}\right\|_{L_2}/r_1 \to 1$ and $\left\|du_{i,j}\right\|_{L_2} (i \neq 1)/r_1 \sim (r_i + r_j)/r_1 \to 0$ at $r_1/r_i \to \infty$. Visually, in this event, the subsets of distances between solutions are manifested as clusters of points, if the distances between solutions $\left\|du_{i,j}\right\|_{L_2}$ are ordered in accordance with their magnitude (a visual illustration is presented below, see Figs. 3, 5).

The maximum of $\left\|du_{i,j}\right\|_{L_2} (i \neq 1)$ (the distance from zero to maximal value in the cluster) is considered as the upper bound of first cluster $\delta_1$ (distances between "accurate" solutions), while for the minimum of $\left\|du_{1,j}\right\|_{L_2}$ we define as a down border of the second cluster $\delta_2$ (distances between "accurate" solutions and most inaccurate one).

The separation of distances between solutions into clusters may be considered as evidence of the existence of solutions with significantly different error norms. The quantitative criterion for applicability of *Theorem 1*, based on dimension of first cluster and the distance between clusters, may be stated as the following heuristic **Criterion 1**:

*If the distance between clusters is greater than the size of the cluster of accurate solutions $\delta_2 - \delta_1 > \delta_1$, then $\left\|\tilde{u} - u^{(i)}\right\|_{L_2} \leq \left\|du_{1,i}\right\|_{L_2}$, where $u^{(i)}$ belongs to the cluster of more accurate solutions and $u^{(1)}$ is the maximally inaccurate solution.*

Let $r_{i,\max}$ be the maximum error norm in the subset of accurate solutions. The *Criterion 1* is based on the assumptions that the dimension of the "accurate" cluster is equal to double maximum error in the cluster ($\delta_1 = 2r_{i,\max}$) and the second cluster belongs to the interval ($r_1 - r_{i,\max}, r_1 + r_{i,\max}$) ($\delta_2 = r_1 - r_{i,\max}$). Both these evaluations are overestimated and correspond to collinear vectors of the error. Under these assumptions the relation of accurate cluster dimension and the distance between clusters ($\delta_2 - \delta_1 > \delta_1$) assumes the form $r_1 - r_{i,\max} > 4r_{i,\max}$. This leads to the relation $r_1 > 5r_{i,\max}$ that involves condition (2) $r_1 > 2r_{i,\max}$ and serves as the justification of *Criterion 1*.

The *Criterion 1* may be rigorous only in the limit of an infinite set of solutions, computed by independent methods. Nevertheless, the numerical check for this criterion confirmation or violation is of interest from the viewpoint of its applicability as heuristics.

### 4. Numerical Tests

The results of the error norm estimation using above mentioned criterion are presented below for several test flows governed by two dimensional unsteady Euler equations.

$$\frac{\partial \rho}{\partial t} + \frac{\partial \left(\rho U^k\right)}{\partial x^k} = 0 ; \tag{4}$$

$$\frac{\partial \left(\rho U^i\right)}{\partial t} + \frac{\partial \left(\rho U^k U^i + P\delta_{ik}\right)}{\partial x^k} = 0 ; \tag{5}$$

$$\frac{\partial (\rho E)}{\partial t} + \frac{\partial \left(\rho U^k h_0\right)}{\partial x^k} = 0 . \tag{6}$$

Here the summation over repeating indexes is assumed, $i,k = 1,2$, $U^1 = U, U^2 = V$ are the velocity components, $h_0 = (U^2 + V^2)/2 + h$, $h = \dfrac{\gamma}{\gamma-1}\dfrac{P}{\rho} = \gamma e$, $e = \dfrac{RT}{\gamma-1}$, $E = \left(e + \dfrac{1}{2}(U^2 + V^2)\right)$ are enthalpies and energies, $P = \rho RT$ is the state equation and $\gamma = C_p / C_v$ is the specific heat ratio.

The single oblique shock wave, the interaction of shock waves of I and VI kinds according to Edney classification [21] were used as the test problems. Only steady-state solutions were considered, so only the spatial discretization error is addressed. Several analytical solutions were constructed for these problems. The shock wave is the main element of these solutions. First, the shock wave angle $\beta$ was computed from the flow deflection angle $\alpha$ using the expression [22]

$$\frac{tg(\beta-\alpha)}{tg\beta} = \left(\frac{(\gamma-1)M_\infty^2 \sin^2\beta + 2}{(\gamma+1)M_\infty^2 \sin^2\beta}\right), \tag{7}$$

which was resolved iteratively. Flow parameters past shock wave $(f_1)$ were computed from parameters $f_\infty$ before shock via Rankine–Hugoniot conditions

$$M_1^2 = \frac{M_\infty^2 + 2/(\gamma-1)}{2\gamma/(\gamma-1)M_\infty^2 \sin^2\beta - 1} + \frac{M_\infty^2 \cos^2\alpha}{1/2(\gamma-1)M_\infty^2 \sin^2\beta + 1}, \tag{8}$$

$$\rho_1/\rho_\infty = 1 + \frac{M_\infty^2 \sin^2\alpha - 1}{1 + (\gamma-1)/2M_\infty^2 \sin^2\alpha}, \tag{9}$$

$$T_1/T_\infty = \{2\gamma \cdot M_\infty^2 \sin^2\alpha - (\gamma-1)\}\frac{2 + M_\infty^2 \sin^2\alpha \cdot (\gamma-1)}{(\gamma+1)^2 M_\infty^2 \sin^2\alpha}. \tag{10}$$

For the single shock wave test these parameters are sufficient for the flow field generation.

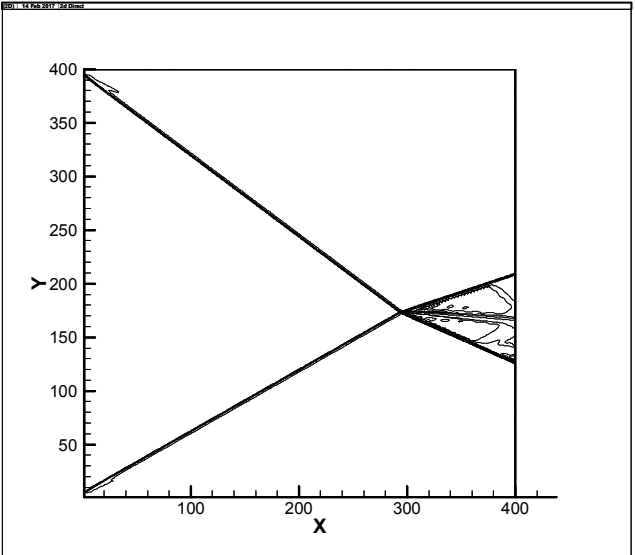

Fig. 1. Edney I flow density isolines. $M = 5$, flow deflection angles $\alpha_1 = 20^o$ and $\alpha_2 = 26^o$.

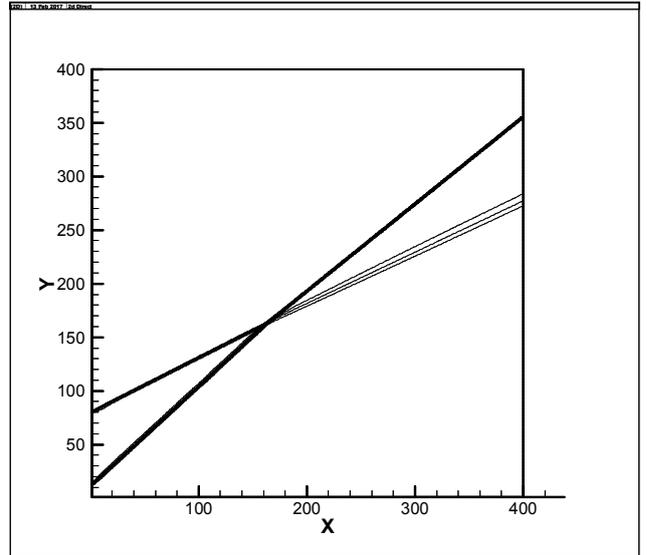

Fig. 2. Edney VI flow density isolines. $M = 4$, Two consecutive flow deflection angles $\alpha_1 = 10^o$, $\alpha_2 = 15^o$..

For the Edney I shock structure (Fig. 1), an additional iteration was used to determine the angles of shock waves past crossing. It was performed by fitting the contact line direction such that the pressures on both sides of the contact line coincide.

For the Edney VI shock interaction (Fig. 2) the flow field was computed using two consequent oblique shocks and the single shock past the point of initial shocks crossing. The contact line and additional weak wave of opposite family [22] emerge at the crossing point. The additional wave (shock or expansion fan) should be computed to equalize pressures on both sides of the contact line. In the present tests, it was the expansion fan (Prandtl-Meyer flow).

The resulting flow field was projected to the computational grid and the result was considered as the "exact" solution.

The flow field contains undisturbed domains (nominal order of error, declared by authors of numerical methods, is expected), shock waves (error order about $n = 1$ [7]), contact discontinuity line (error order about $n = 1/2$, [6]). As a result, one may hope to obtain a nontrivial error composed of components with different error orders.

The numerical computations were performed for Mach numbers $M = 3, 4, 5$, flow deflection angles range of $\alpha = 10 - 30^o$ and $C_p / C_v = 1.4$. For example, Fig. 1 presents the density isolines for Edney I flow structure ($M = 5$, flow deflection angles $\alpha_1 = 20^o$ and $\alpha_2 = 26^o$). The crossing shock waves and contact discontinuity line, engendered at the shocks crossing point, are the main elements of this flow structure. Fig. 2 presents the density distribution for Edney VI flow structure ($M = 4$, two consequent flow deflection angles $\alpha_1 = 10^o$, $\alpha_2 = 15^o$). The flow is determined by the merging shock waves, the contact line and the expansion fan.

The paper contains an analysis of the ensemble of computations performed by methods listed below.

The first order scheme by Courant Isaacson Rees (CIR) [23] is referred to as $S1$.

The second order scheme using the MUSCL method [24] and algorithm by [25] at cell boundaries is denoted as $S2$.

Second order TVD scheme of relaxation type by [26] is referred to as $S2TVD$.

Third order modified Chakravarthy-Osher scheme [27, 28] is marked as $S3$.

Fourth order scheme by [29] is referred to as $S4$.

Computations were performed on uniform grids.

The total number of the test configurations (Mach numbers (M=3, 4, 5), grid sizes ($100 \times 100$, $400 \times 400$ nodes), flow structure (single shock, Edney I and Edney VI), 6 additional tests for Edney I and Edney VI at different deflection angles (M=4, $400 \times 400$ nodes)) was equal to 24. For every configuration, the flow field was computed using $S1, S2, S3, S4$ and S2TVD solvers, respectively.

The vector of solution of Eq. (4-6) contains four components $u^{(i)} = \{\rho^{(i)}, U^{(i)}, V^{(i)}, e^{(i)}\}$. The distance between solutions was calculated using the following norm

$$\left\| u^{(i)} - u^{(k)} \right\|_{L_2} = \left( \frac{1}{N} \sum_{j=1}^{N} \left( (\rho^{(i)} - \rho^{(k)})_j^2 + (U^{(i)} - U^{(k)})_j^2 + (V^{(i)} - V^{(k)})_j^2 + (e^{(i)} - e^{(k)})_j^2 \right) \right)^{1/2}. \quad (11)$$

It should be noted that methods *S1, S2, S3, S4* (1, 2, 3 and 4 nominal (declared) truncation orders) demonstrates the global order of convergence a bit below $n = 1/2$ in norm $L_2$. The method S2TVD (nominal order 2) demonstrates the global error order of about $n \sim 3/4$.

In numerical tests, we first check *Criterion 1* and, second, we verify the error norm estimation. We consider the $u^{(k)}$ error norm estimation to be successful, if the error estimate $\left\| u^{(i)} - u^{(k)} \right\|_{L_2}$ is greater than the true error norm $\left\| u^{(k)} - \tilde{u} \right\|_{L_2}$ ($\tilde{u}$ is the analytical solution, $u^{(1)}$ - "inaccurate" solution).

The comparison with the analytical solution allows us to conclude that scheme $S1$ (as "inaccurate") and schemes $S2$, $S3$ and $S4$ (as "accurate") enable one to find the upper bound of the error norm, if the *Criterion 1* is satisfied.

Second order $S2TVD$ scheme [26] from the standpoint of error norm is close to the first order scheme $S1$ for $100 \times 100$ grid and to the high order schemes for the grid of $400 \times 400$ nodes. The calculations on the grid $100 \times 100$ demonstrated the formation of clusters with "inaccurate" scheme $S2TVD$ and a successful error estimation. However, the scheme $S2TVD$ on the grid $400 \times 400$ does not form clusters. Paradoxically, the reason for this failure is due to the relatively rapid convergence of $S2TVD$ in comparison with schemes $S2, S3, S4$. As a result, the scheme $S2TVD$ on the grid $400 \times 400$ is close to "accurate" schemes $S2, S3, S4$.

The comparison of schemes $S2, S3, S4$ ($\left\| u^{(2)} - u^{(4)} \right\|_{L_2}$, $\left\| u^{(3)} - u^{(4)} \right\|_{L_2}$, $\left\| u^{(3)} - u^{(2)} \right\|_{L_2}$) does not manifest as splitting into clusters. Similarly, ensembles, containing the pair $S2TVD, S1$, fail to demonstrate splitting into clusters.

For all tests, if the *Criterion 1* is not satisfied (there are no clusters, or distance between them is less the dimension of the cluster of "accurate" solutions) the error norm estimation fails.

The numerical tests for the single oblique shock demonstrate the feasibility for the error norm estimation if the *Criterion 1* is satisfied. However, the set of distances between solutions splits into clusters in about half of tests, more frequently for finer meshes.

For Edney-I shock interaction (Fig. 1), the set of distances between solutions also splits into clusters in about half of the tests without dependence on the mesh size. However, for the distance between clusters, which approximately equals the dimension of the cluster, the error estimation may fail. The worst result over all tests, was obtained in calculations for $M = 5$ and flow deflection angles $\alpha_1 = 20^o$, $\alpha_2 = 26^o$, and is presented in Figs. 3 and 4 ($400 \times 400$ nodes).

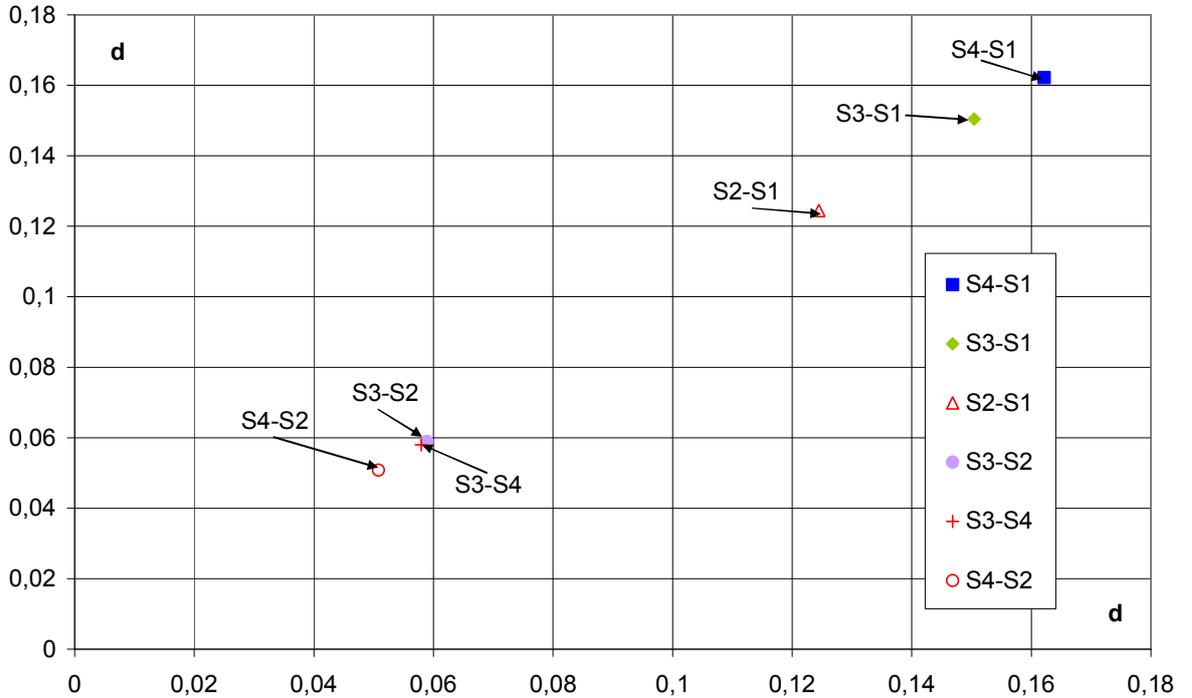

Fig. 3. Clusters for Edney-I flow ($400 \times 400$ nodes). The set of distances between numerical solutions $d = \left\| u^{(i)} - u^{(k)} \right\|_{L_2}$ is marked as $Si - Sk$. The norm is laid out along both axes.

It should be noted that the data under the consideration are bulky, so for ease of visualization, the norm of error is laid out along both axes in Figs. 3-8, even though the data are one dimensional. The norms $d = \left\| u^{(i)} - u^{(k)} \right\|_{L_2}$ are marked as $Si - Sk$.

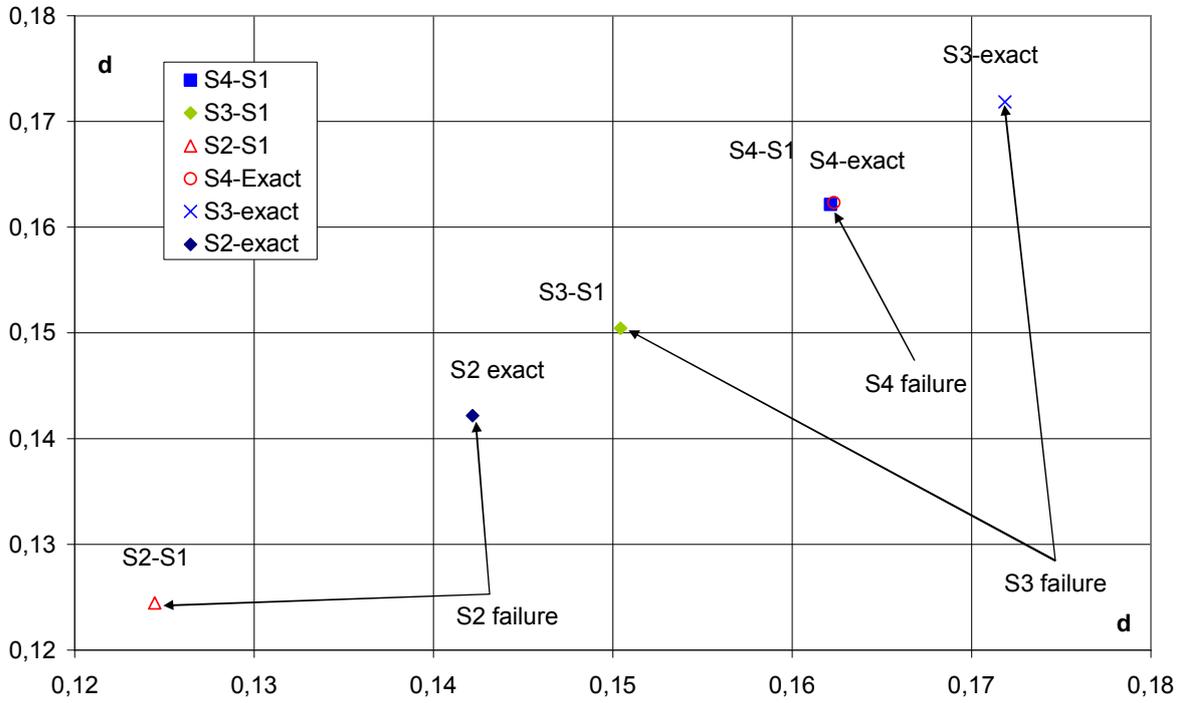

Fig. 4. The error norm estimation (Edney-I ($400 \times 400$ nodes)). The set of distances between numerical solutions $d = \left\| u^{(i)} - u^{(1)} \right\|_{L_2}$ is marked as $Si - S1$ and the set of distances between numerical and analytical ("exact") solutions $d = \left\| u^{(i)} - \widetilde{u} \right\|_{L_2}$ is marked as $Si - exact$. The norm is laid out along both axes.

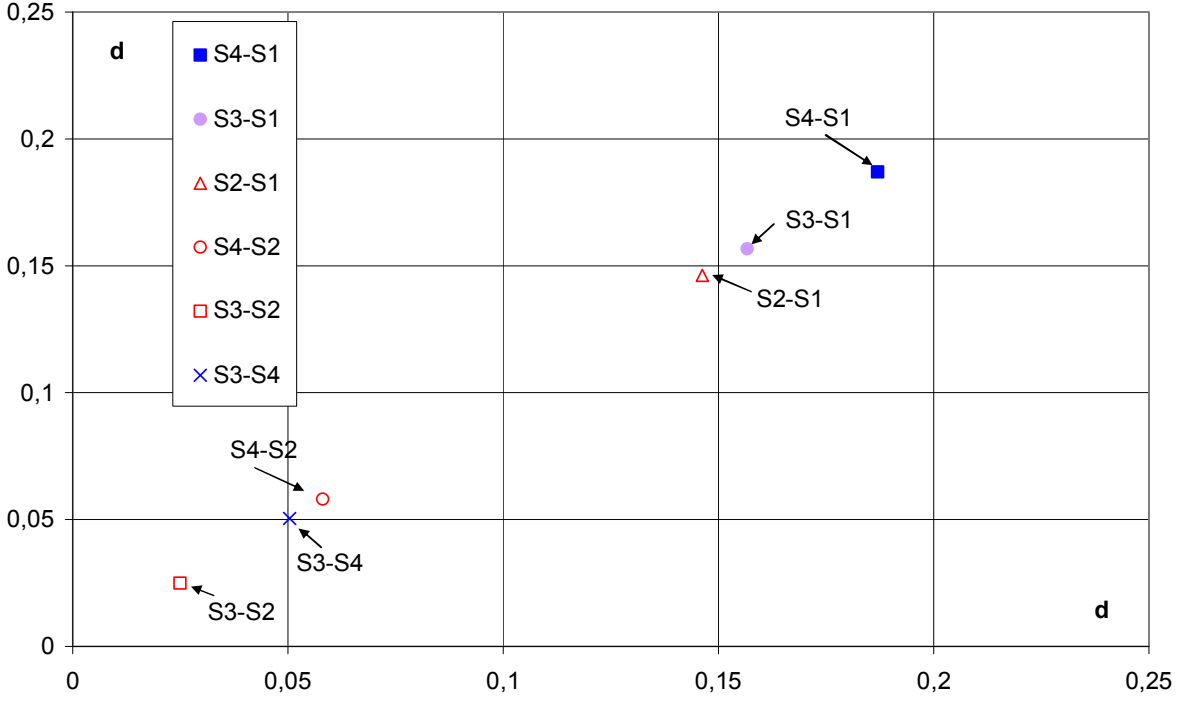

Fig. 5. Clusters for Edney-VI flow ($(100\times100$ nodes).). The set of distances between numerical solutions $d = \left\| u^{(i)} - u^{(k)} \right\|_{L_2}$ is marked as $Si - Sk$. The norm is laid out along both axes.

The maximum relative magnitude of capture condition violation $\left\| \tilde{u} - u^{(2)} \right\|_{L_2} - \left\| du_{1,2} \right\|_{L_2} = \delta_{break}$ is about $\dfrac{\delta_{break}}{\left\| du_{1,2} \right\|_{L_2}} \approx 0.15$. This demonstrates the approximate nature of *Criterion 1,* however, the magnitude of the capture condition violation is not too great.

For Edney-VI shock interaction (Fig. 2) the set of distances between solutions also splits into clusters in about half of numerical tests irrespective of the grid size. However, for the distance between clusters, which approximately equals the dimension of the cluster, the error estimation always performs correctly. The standard result, obtained in calculations for $M = 4$ and flow deflection angles $\alpha_1 = 10^o$, $\alpha_2 = 15^o$, is presented in Fig. 5, 6 ($100\times100$ nodes).

Fig. 5 demonstrates the collection of distances between numerical solutions $\left\| du_{i,k} \right\|_{L_2} = \left\| u^{(i)} - u^{(k)} \right\|_{L_2}$ to break into two clusters, one of them being related to the "inaccurate" scheme $S1$. It enables the successful estimate of the error norm, Fig. 6. It should be noted that the distance between clusters in Fig. 5 is greater than such cluster distance in Fig. 3. Figs. 7, 8 present analogical data for the grid containing $400\times400$ nodes.

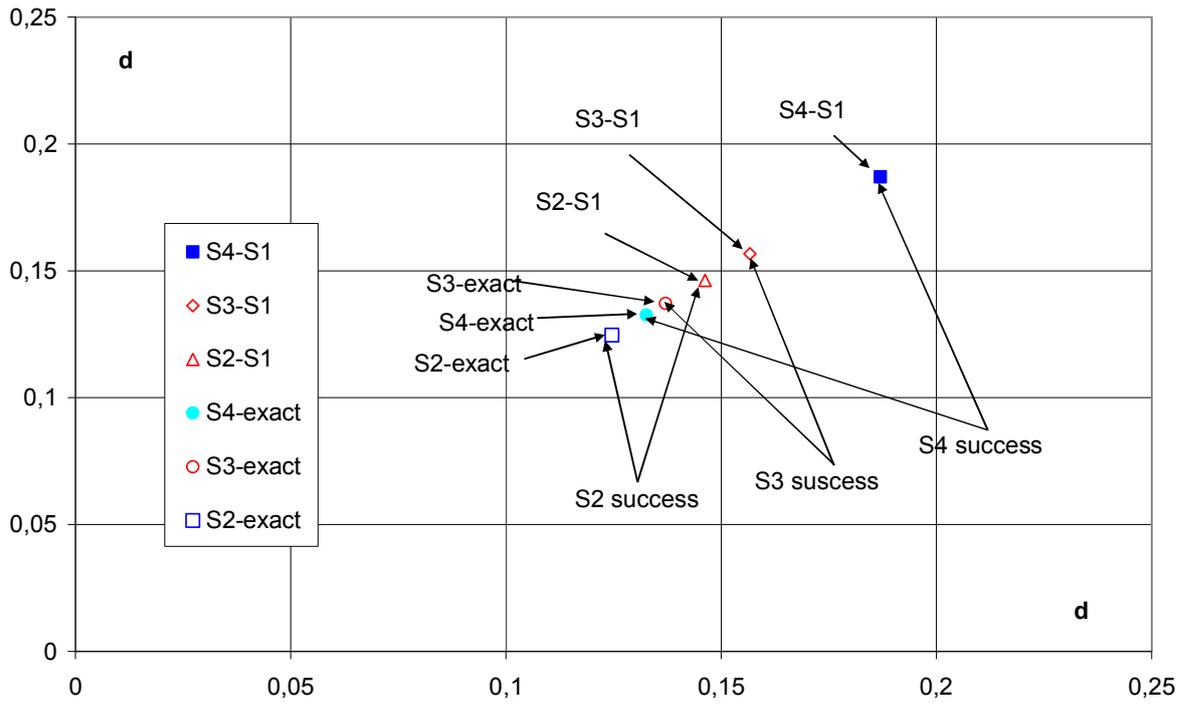

Fig. 6. The error norm estimation (Edney-VI, ($100 \times 100$ nodes)). The set of distances between numerical solutions $d = \left\| u^{(i)} - u^{(1)} \right\|_{L_2}$ is marked as $Si - S1$ and the set of distances between numerical and analytical ("exact") solutions $d = \left\| u^{(i)} - \tilde{u} \right\|_{L_2}$ is marked as $Si - exact$. The norm is laid out along both axes.

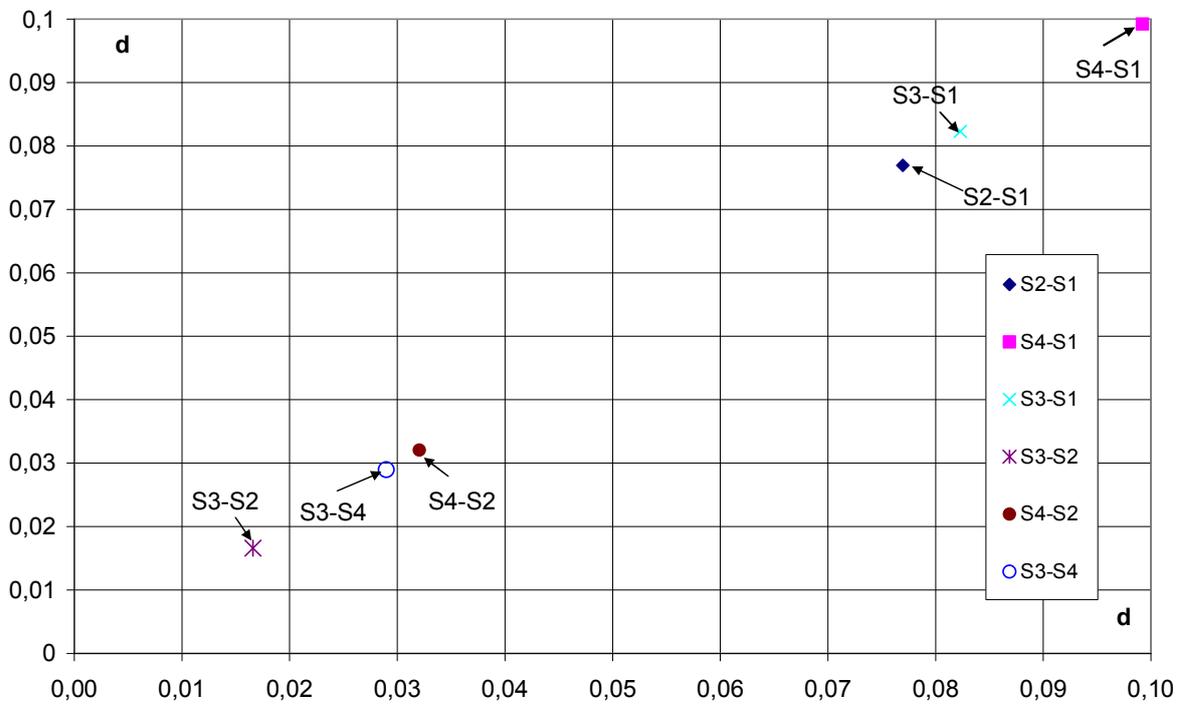

Fig. 7. Clusters for Edney-VI flow ($400 \times 400$ nodes). The set of distances between numerical solutions $d = \left\| u^{(i)} - u^{(k)} \right\|_{L_2}$ is marked as $Si - Sk$. The norm is laid out along both axes.

The results demonstrate the successful estimation of error and the convergence order about $1/2$ for all tested methods without dependence on the formal order of approximation. This result stresses the differences between current approach and "p-refinement" (p-FEM) widely used in the domain of finite elements [30,31].

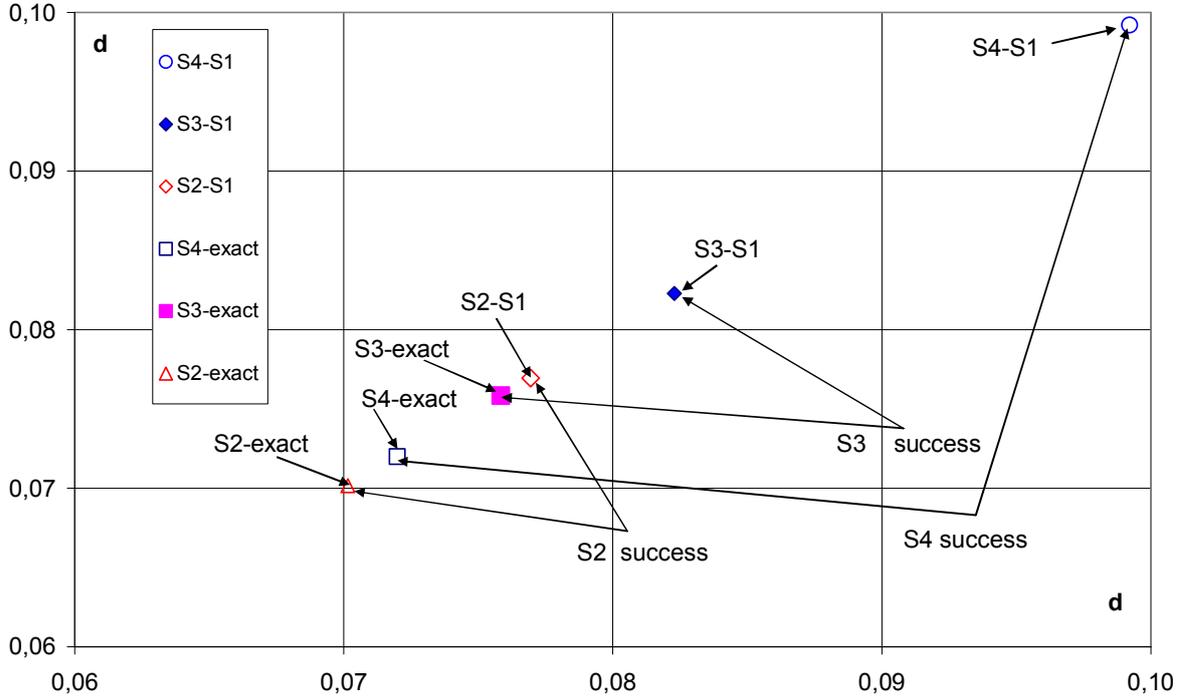

Fig. 8. The error norm estimation (Edney-VI, ($400 \times 400$ nodes)). The set of distances between numerical solutions $d = \left\| u^{(i)} - u^{(1)} \right\|_{L_2}$ is marked as $Si - S1$ and the set of distances between numerical and analytical ("exact") solutions $d = \left\| u^{(i)} - \tilde{u} \right\|_{L_2}$ is marked as $Si - exact$. The norm is laid out along both axes.

Thus, for the estimation of error norm upper bound, one should have a priori information regarding error rating (*Theorem 1*) or the ensemble of minimum three solutions with the distances between them split into two clusters. The distance between clusters should be greater than the dimension of the cluster of distances between more accurate solutions (*Criterion 1*).

Two thirds of the numerical tests for two dimensional supersonic inviscid flows confirm the estimation (3) $\left\| \tilde{u} - u^{(i)} \right\|_{L_2} \leq \left\| du_{1,i} \right\|_{L_2}$, if the heuristic *Criterion 1* is satisfied. In one third of tests, the estimation (3) failed, however, the maximal observed violation of expression (3) is found to be about 15%.

The relation of errors obtained in the paper is not necessarily attributed to properties of the considered schemes. In a strict sense, it may be caused by the imperfections of numerical realization performed by the paper authors. So, the authors do not pretend to provide a definitive assessment of the methods considered. Our purpose is rather to verify the single-grid error estimator based on the numerical results obtained by the solvers (algorithms realizations) of different accuracy.

## 5. Discussion

The determination, whether the scheme is "accurate" or "inaccurate" one, has the asymptotic sense in *a priori* analysis. It may also be performed by the comparison with the limited number of analytic solutions. The above results demonstrate the feasibility to distinguish "accurate" and "inaccurate" numerical solutions in the sense of discretization error norm rating. For example, the distribution of distances between solutions $\left\|du_{i,j}\right\|_{L_2}$, presented in Figs. 3,5,7, demonstrates two clusters corresponding to "accurate" and "inaccurate" schemes. This provides the possibility of finding the error norm only from observable values $\left\|du_{i,j}\right\|_{L_2}$ (without *a priori* information on errors ranging), that is confirmed by Figs. 6,8. The results presented in Figs. 6,8 demonstrate the standard quality of the error norm estimation obtained in most tests, if conditions by *Criterion 1* are satisfied. The violation of condition $\left\|\tilde{u} - u^{(2)}\right\|_{L_2} \leq \left\|du_{1,2}\right\|_{L_2}$ above 15% (Fig. 4) is not detected in tests.

At first glance, the present approach is similar to the "p-refinement", widely used in the domain of finite elements [30]. However, "p-refinement" estimates the error of the *lower order solution (less precise)* by difference between it and the high order solution. Herein, we majorize the error of *more precise* solution by the difference between solutions under specific CFD conditions (shock waves and contact lines), when schemes of any formal order of approximation have the same real order of convergence. Some works discuss the analogue of Richardson extrapolation [31], which utilizes three finite element solutions with consequent orders of accuracy, that is of some analogy with our technique. However, algorithm [31] is based on specific asymptotic of energy norms and is not related with the triangle inequality and formation of clusters.

The above considered single-grid discretization error estimator operates with the total error including the discretization error in flow field, initial and boundary condition error and round-off errors. It is used in a postprocessor mode like the Richardson extrapolation. However, it does not require any mesh refinement and may be used away from the asymptotic range.

The dependence on the set of numerical methods and analyzed solution is the drawback of the ensemble based estimator. The same set of methods may provide a segregation into clusters for one flow pattern and may not provide it for another. So, this approach cannot replace the mesh refinement and is only aiming to supplement it by a non-expensive algorithm.

### *6. Conclusions*

It is feasible to estimate the discretization error norm using a collection of numerical solutions obtained on the same grid by solvers of different orders of approximation.

If the collection of solutions is split into separate clusters, corresponding to "accurate" and "inaccurate" schemes and the distance between clusters is greater than the dimension of the "accurate" cluster, the norm of the error of the more accurate solution is majorized by the norm of the solutions difference.

Numerical tests demonstrated the applicability (with a reasonable tolerance) of this heuristic rule in $L_2$ for two dimensional supersonic problems (containing shocks and contact discontinuities) governed by the Euler equations.

The above considered single-grid discretization error estimator may be constructed using an ensemble of numerical solutions obtained by different solvers of various orders of accuracy. It is used in a non-intrusive postprocessor mode and does not require mesh refinement.

### *Acknowledgement*

A.K. Alekseev and A.E. Bondarev acknowledge the partial support by grants of RFBR № 16-01-00553A and 17-01-444A.